# CIoTA: Collaborative IoT Anomaly Detection via Blockchain

Tomer Golomb, Yisroel Mirsky and Yuval Elovici
Ben-Gurion University of the Negev
{golombt, yisroel}@post.bgu.ac.il, {elovici}@bgu.ac.il

*Abstract*—Due to their rapid growth and deployment, Internet of things (IoT) devices have become a central aspect of our daily lives. However, they tend to have many vulnerabilities which can be exploited by an attacker. Unsupervised techniques, such as anomaly detection, can help us secure the IoT devices. However, an anomaly detection model must be trained for a long time in order to capture all benign behaviors. This approach is vulnerable to adversarial attacks since all observations are assumed to be benign while training the anomaly detection model.

In this paper, we propose CIoTA, a lightweight framework that utilizes the blockchain concept to perform distributed and collaborative anomaly detection for devices with limited resources. CIoTA uses blockchain to incrementally update a trusted anomaly detection model via self-attestation and consensus among IoT devices. We evaluate CIoTA on our own distributed IoT simulation platform, which consists of 48 Raspberry Pis, to demonstrate CIoTA's ability to enhance the security of each device and the security of the network as a whole.

## I. INTRODUCTION

The Internet of Things (IoT) is the next evolution of the Internet [12]. Leading IoT experts believe that by 2020 there will be more than 50 billion devices connected to the Internet which will offer a variety of applications and services for both daily and critical uses [8]. The vision for these IoT devices is that they will autonomously communicate with one another to improve services and our daily lives.

Like any new architecture or technology, the IoT improves our lives but introduces disruptive elements as well. One known issue associated with IoT devices is that they tend to have vulnerabilities, which in some cases go unpatched by the manufacturer. An attacker can exploit these vulnerabilities for nefarious purposes [20]. Since IoT devices have been integrated into both daily and critical applications, their security is a significant concern.

One solution for protecting an IoT device is to install an intrusion detection system (IDS) [9]. If the IDS is anomaly-based, it has the potential to detect new and emerging threats that target the device. However, there are two fundamental challenges with anomaly-based IDSs:

1) **Adversarial Attacks** During the training phase, all observations are considered benign and are used to train a model which captures the device's normal behavior. After the training phase, the model enters an execution phase where it is used to detect when newly observed behaviors deviate from the norm. Therefore, an attacker who exploits a device before the execution phase can evade detection entirely.
2) **False Positive** It is likely that false positives will occur if an anomaly detection model is trained on the observations of just a few devices. This is because some legitimate behaviors are rare and event-based, and therefore may not appear in the training data. For example, the motion detection logic of a smart camera or the response generated by a smoke detector sensing a fire.

However, consider the following scenario. Assume that all IoT devices of the same type simultaneously begin training their own anomaly detection model, based on their own locally observed behaviors. In this scenario, it is unlikely that the majority of these IoT devices would be exploited before they completes their training phases.

Using this underlying assumption, we present CIoTA (pronounced as *syota*): a lightweight, scalable framework which utilizes the blockchain concept to perform distributed and collaborative anomaly detection on resource limited devices, such as IoT devices.

The blockchain is a protocol for maintaining a distributed ledger. The ledger is a chain of blocks which is collectively agreed upon by the majority of participants in the network [23]. Each block is accepted into the chain if it can achieve consensus represented as the satisfaction of specific criteria (e.g., the proof of work criterion in bitcoin [15]). In the context of a device from a specific model, CIoTA uses the blockchain protocol to collaborate among devices of the same type to create a trusted anomaly detection model. This is accomplished by merging locally trained models into a single global model. A block is a set of locally trained models from different devices. Each device validates the integrity of a block in progress by merging the models and by performing self-attestation. When the block reaches a size limit, it is closed and a new block begins. One closed block in the chain represents a trusted model which has been validated by the majority of devices in the system, and is therefore ready to be used (executed on-site). However, one may wish to track a chain of models, since a stronger model can be created by combining multiple blocks together.

The anomaly detection model used by CIoTA is an extensible Markov model (EMM). The EMM tracks a program's regular memory jump sequences and can be incrementally updated and merged with other models. To introduce new benign concepts into the merged model, we accept new states to the EMM only when there is a consensus among the models.

We evaluate CIoTA on our own IoT emulation platform, involving 48 Raspberry Pis. Our evaluation demonstrates CIoTA's capability in detecting local attacks, and CIoTA's resistance against adversarial attacks (on CIoTA itself). To encourage further research and development, the reader may download our data and source code from GitHub[1].

## II. RELATED WORKS

One proposed solution for protecting IoT devices is to deploy an anomaly-based IDS on the device itself [3], [17], [19], [24]. However, these solutions neglect the facts that (1) the training phase is sensitive to adversarial attacks, and (2) rare benign activities (which do not appear in the initial training data) can cause false positives. Furthermore, some of these solutions [1], [11], [18] require a centralized server and therefore do not scale well to large numbers of IoT devices. Unlike other IDS solutions, CIoTA continuously learns, is robust to adversarial attacks, and is highly scalable.

Another solution, proposed in the literature, is to deploy a network-based IDS (NIDS) [5], [24]. However, in many cases, IoT devices cannot be topologically placed behind a NIDS, making these solutions impractical. CIoTA does not depend on the IoT device's network topology, and therefore can be applied in a general environment.

In [10], [16] the authors propose deploying static analysis tools on the IoT devices. However, this approach requires that (1) the device maintains a database of virus signatures and (2) that experts continuously update this database. CIoTA is anomaly-based and therefore updates itself automatically without human intervention.

Other researchers have tried to avoid the issue of training altogether, by using a trust anchor, such as an IoT device's functional relationship. In [13], the authors propose executing every computation twice across different IoT devices and then compare the results to detect deviations (infected devices). However, this method was only designed to secure against specific types of attacks, whereas CIoTA is generic.

Other trust anchors are the Trusted Platform Module (TPM) [14] and Trusted Execution Environment (TEE) [25]. ARM's TrustZone [22] is a TEE implemented in the hardware, providing a one-way separation between two worlds: "unsecured" and "secured". C-FLAT [1] utilizes the TrustZone for attesting the IoT device's control-flow behavior against a simulation run in parallel on a central server. Although an application's control-flow can be used to detect a vast range of code execution attacks, C-FLAT is limited to specific IoT devices which (1) do not execute code continuously or (2) devices whose behavior is not affected by external sensory events (e.g., smart cameras). CIoTA analyzes control-flow behavior to detect abnormalities on site and therefore does not have the limitations of C-FLAT.

[1] https://git.io/vAIvd

## III. TECHNICAL BACKGROUND & NOTATION

### A. Terminology

**Model** An anomaly detection model $M$ that supports (1) the calculation of a distance between models, and (2) combining (merging) several models of the same type together. We denote a model which is currently deployed on a local device as $M_L$.

**Verified Model** Let $d(M_i, M_j)$ be the distance between models $M_i$ and $M_j$. A model $M_v$ is said to be verified by a device if $d(M_L, M_v) < \alpha$, where $\alpha$ is a user given parameter of the system.

**Combined Model** A model created by merging a set of models together. The combined model only contains elements which are present in at least $p_c$ percent of the models in the set.

**Report** A record consisting of a model $m$ and a cryptographic signature $S(m, seed)$ that uses a private key on the model and a random seed.

**Block** A list of reports on the same seed from different devices which have precisely $L$ entries. If a block has less than $L$ records, then it is referred to as a *partial block* or as a *block in progress*. A device would contribute to expanding a *partial block* only if it deem the block content as verified.

**Trusted Model** The combined model is resulting by merging the models in the reports of the most recently closed block. Since reports come from different devices (verify by their signatures), and the reports are verified at each extension of the block, the combined model is considered to have collaborative trust.

**Chain** A blockchain consisting of CIoTA's blocks for a specific type of IoT device. The block's acceptance criterion is for the block to include precisely $L$ entries, while the criterion for accepting a *partial block* is for the *combined model* which results from merging the models, to be a *verified model*. The length of a chain is the total number of completed blocks in that chain. Finally, a chain may have at most one *partial block* appended to the end of the chain.

**Agent** A program that runs on an IoT device which is responsible for (1) training and executing the local model $M_L$, (2) downloading broadcasted chains to replace $M_L$ and the locally stored chain, and (3) periodically broadcasting the locally stored chain, with the latest $M_L$ as a report in the *partial block*. An agent only replaces his/her locally stored chain if the downloaded chain is longer than the existing chain. Furthermore, an agent only uses the downloaded chain's *partial block* if the agent can attest that the combined model, resulting from the *partial block*, is a *verified model*.

### B. Extensible Markov Model (EMM)

A Markov chain (MC) model is a *memory-less process*, i.e., a process where the probability of transition at time $t$ only depends on the state at time $t$ and not on any of the states leading up that state. Typically, an MC is represented as an adjacent matrix $M$, such that $M_{ij}$ stores the probability of transitioning from state $i$ to state $j$ at any given time $t$. Formally, if $X_t$ is the random variable representing the state

at time $t$, then
$$M_{ij} = Pr(X_{t+1} = j | X_t = i) \tag{1}$$

An EMM [4] is the incremental version of the MC. Let $N = [n_{ij}]$ be the frequency matrix, such that $n_{ij}$ is the number of transitions which have occurred from state $i$ to state $j$. From here, the MC can be obtained by

$$M = [M_{ij}] = \left[\frac{n_{ij}}{n_i}\right] \tag{2}$$

where $n_i = \sum_j n_{i,j}$ is the total number of outgoing transitions observed by state $i$.

In CIoTA, an MC captures an application's control-flow asynchronously by using Hardware counters. To restrict the size of the MC, a state in the MC is a region (an address range) in the application's memory. Finally, a transition is the probability that the program will jump from region $i$ to region $j$. To update the EMM, an agent increments $n_{ij}$ asynchronously, whenever a control-flow event occurs. Anomalies are detected when the observed $M_{ij}$ is less than a user defined probability $p_{thr}$.

Let $\mathbf{N}$ be a set of EMM models, and let $\mathbf{N}_{ij}^{(k)}$ be the element $N_{ij}$ in the $k$-th model in $\mathbf{N}$. To create a combined model from the set $\mathbf{N}$, Algorithm 1 is performed, where $p_c$ is the minimum consensus percent requirement for distinguishing between attacks and rare behaviors. After forming the MC model $M$ from $N$ using (2), an agent can attest that $M$ is a verified model using Algorithm 2, where $\alpha \in [0, 1]$ is the verification distance cutoff, provided by the user.

## IV. THE CIoTA BLOCKCHAIN

### A. Overview

As a manner of analogy, let's assume that there is an agency called CIoTA which has many agents, each of which is in enemy's territory. Their mission is to continuously (1) detect malicious acts, and (2) gather Intel about what is happening and periodically share a collection of Intel (as rumors) with other agents. Since an agent is in the enemy's territory, the agent may receive false Intel which introduces noise into the rumors conveyed to the other agents. Therefore, an agent will only trust a rumor (some Intel) if the agent knows that at least $p_c$ other agents have heard the same rumor. Finally, an agent accepts the most recent set of $L$ trusted rumors as the latest description of the territory.

This scenario is implemented in the CIoTA framework as follows. Each IoT device has an agent which maintains a local model $M_L$ that is used to detect malicious behaviors in a particular application. An agent records new Intel by updating $M_L$ with observations on the application's behavior. An agent shares its Intel $(M_L)^2$, in the form of a rumor, by adding $M_L$ to the chain's *partial block*, and then sending the chain to neighboring agents in the network. Other agents will only accept this *partial block* if it is longer than their *partial block*, and if they can attest that it is safe (by verifying it against their own local model). Thus, the *partial block* only grows if the majority of agents have verified that it contains a safe model. Once the *partial block* have $L$ reports, it is closed as a completed block. Therefore, an agent receives the latest intelligence from its fellow agents by replacing $M_L$ with the combined model contained within the newest closed block. Finally, to ensure that the rumors have indeed come from a specific agent, record (rumors) are signed by a private key and verified with the respective public key.

[2]An MC is converted during sending and reception to match the device state (ASLR).

**Algorithm 1** The algorithm for combining a set of EMMs.
**function** COMBINE($\mathbf{N}, p_c$)
    $N \leftarrow$ empty_EMM()    ▷ initialize empty freq. matrix
    **for** $n_{ij} \in N$ **do**
        $C \leftarrow 0$    ▷ init the consensus counter
        **for** $k \in 1 : |\mathbf{N}|$ **do**
            $n_{ij} \leftarrow n_{ij} + \mathbf{N}_{ij}^{(k)}$
            **if** $\mathbf{N}_{ij}^{(k)} > 0$ **then**
                $C++$
        **if** $\frac{C}{|\mathbf{N}|} \leq p_c$ **then**    ▷ no consensus on element $ij$
            $n_{ij} \leftarrow 0$
    **return** $N$

**Algorithm 2** The algorithm for attesting that the MC $M$ is a verified model with respect to the local model $M_L$
**function** VERIFY$_{M_L}(M, \alpha)$
    $l_1 \leftarrow 0$    ▷ init linear norm
    **for** $M_{ij} \in M$ **do**
        $l_1 \leftarrow l_1 + |M_{ij} - M_{L,ij}|$
    $n_{states} \leftarrow \dim(M_L)$
    **if** $\frac{l_1}{n_{states}} \leq \alpha$ **then**    ▷ the linear distance is acceptable
        **return** True
    **else**    ▷ the linear distance is too large
        **return** False

### B. An Agent's Processes

The following describes an agent's three parallel processes, as illustrated in the flowchart in Fig. 1.

*1) Gather Intelligence:* For simplicity, each IoT device starts by creating its local model $M_L$ and then goes on to continuously collect Intel about its territory (local application), and continuously detect anomalies in the collected Intel using $M_L$. If the $M_L$ does not raise an alert, the device updates $M_L$ to support the new Intel. An exception is when $M_L$ is a new model, as opposed to a combined model downloaded from the chain. In this case, $M_L$ is given a brief period of training from all Intel unconditionally. If infected, this short period will not affect the other agents, since rumors (broadcasted models) are only accepted if verified by the majority of agents.

*2) Receive Intelligence:* When the local agent $A$ receives chain $C_B$ from agent $B$, $A$ checks if $C_B$ is (1) longer than $C_A$ and (2) contains valid signatures. If these conditions hold, then the blocks in $C_A$ are replaced by those in $C_B$, and the last completed block is used to form a combined model (Algorithm 1) which replaces $M_L$. If the *partial block* in $C_B$ is longer than the *partial block* in $C_A$ and the combined model from the *partial block* passes the validation test in Algorithm 2, then $C_A$ is updated with the *partial block* from $C_B$.

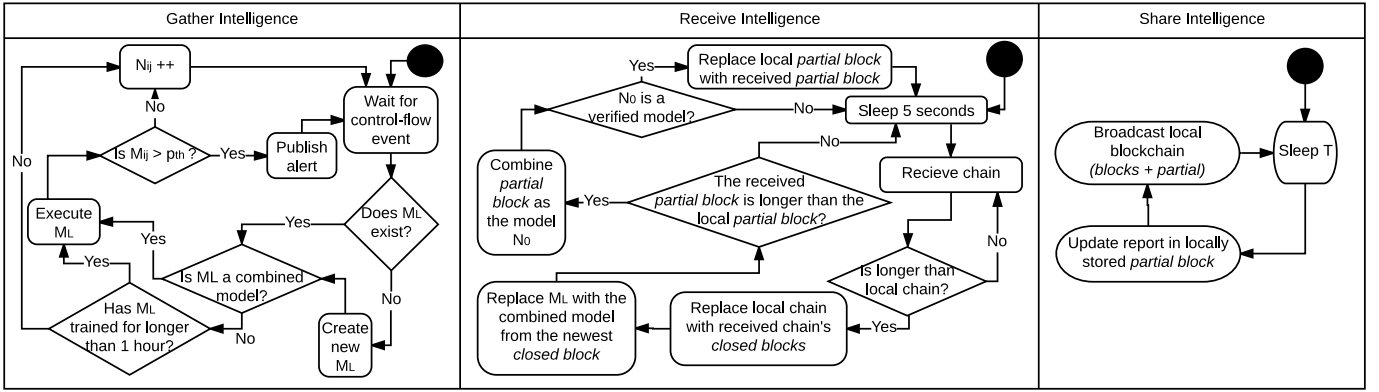

Fig. 1. A flow-chart of an agent's three asynchronous processes.

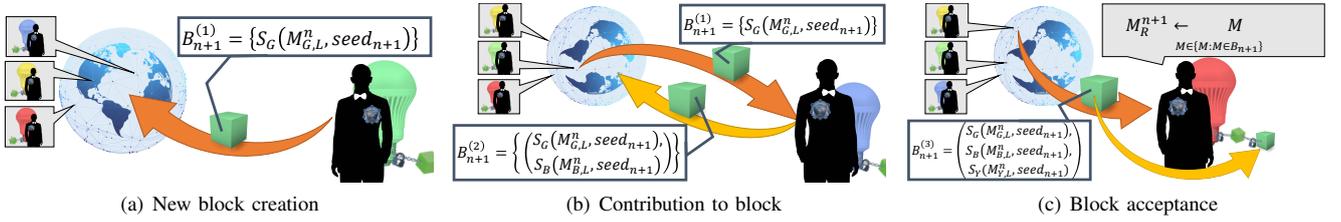

(a) New block creation  (b) Contribution to block  (c) Block acceptance

Fig. 2. CIoTA High level protocol

*3) Share Intelligence:* Periodically every $T$ time elapses, an agent updates its report in the locally stored chain's *partial block*. The agent then broadcasts the entire locally stored chain to all neighboring agents.

To further illustrate the system, let's assume CIoTA has been installed on a set of IoT smart bulbs, and the maximum number of reports in a block ($L$) is 3. Each device maintains a list of the last $N$ blocks, $B_{n-N}$ to $B_n$, to limit the amount of memory consumed by the locally stored chain. If there is no *partial block*, a bulb can create a new *partial block* or add/update the existing *partial block*. Fig. 2(a) illustrates the green bulb's attempt to create $B_{n+1}$ with the device's report on its model, $M_{G,L}$ and randomly generated $seed_{n+1}$. In Fig. 2(b) the blue bulb contributes to $B_{n+1}$ by expanding $B_{n+1}$ with its report, which is a signature on $M_{B,L}$ and $seed_{n+1}$. The yellow bulb completes the block by adding its report to the block, causing all of the agents to update their model with a combined model based on all of the models in $B_{n+1}$. (Fig. 2(c)).

### C. The Security of the Chain

CIoTA's central assumption is that an attacker cannot exploit a large number of devices within a short period, while simultaneously evading $M_L$'s detection. CIoTA also assumes that the reports in a block came from the reporting device; otherwise, an attacker could merely broadcast fake chains. To validate the source, CIoTA can use either (1) PKI (public-key infrastructure [2]), which allows identification and authentication of every device, or (2) a shared secret (symmetric key) among the agents, which prevents an outsider's interference in the protocol.

PKI management is cumbersome and introduces new dangers [7], and therefore we implement CIoTA using symmetric key encryption inside the IoT devices' TrustZone. The TrustZone is a safe house on a device which has access to untrusted territory within the device. In this setup, the agent, along with a symmetric encryption key (provided by the admin), are located in the TrustZone. By doing so, the agent will be able to securely communicate with other agents, while avoiding the issues of PKI. However, an incorrect implementation can lead to severe vulnerabilities that affect the whole network.

## V. ATTACKS AGAINST CIoTA

In this section, we discuss possible attack vectors against CIoTA and their implications.

### A. Attacks Against the Agent

The goal of an attacker is to execute his/her own persistent logic on the IoT device. An attacker might, during the attack, negatively affect $M_L$ either before or after the training phase. In both cases the effect will occur before the agent shares it with other agents via the *partial block*. However, in this case, the majority of agents will either (1) reject the *partial block* because the verification Algorithm (2) will fail, or (2) omit the additional behaviors (new transitions) while forming the combined model in Algorithm (1). Moreover, if the agent is in the TrustZone, then $M_L$ will be replaced when the next block will closed, and then the malware will be detected.

Another possibility is that the attacker will try and sabotage the agent. However, since the agent is installed at a higher level than the infected application (e.g., either kernel or TrustZone), accessing the agent's memory address will require additional exploits. This requirement creates a *catch-22* since the attacker needs to perform additional actions (by exploiting the device further) to access the agent, yet doing so creates anomalies in the control-flow which in turn raises alerts.

## B. Attacks Against the CIoTA Chain

An attacker can alter the anomaly detection model by creating the same alterations to at least $p_c$ percent of the models inside the *partial block*, and then broadcast the chain. However, this is challenging because (1) the deployment may be widespread geographically, and (2) the attacker must obtain a large number of private keys when employing PKI or extract the symmetric key from the TrustZone. We note that to ensure a strong level of security, the values of $p_c$ and $L$ should be relatively large, and PKI should be used if possible.

## C. Detection Policies

While detection is a powerful tool for security, without action is meaningless. Therefore, we suggest that CIoTA's agents apply one of the following policies after detecting an attack: (1) send alerts to a control server, (2) restart the infected application, or (3) stop the device entirely via the kernel. Selecting a policy depends on the IoT device's application and how critical its operations are.

## VI. EVALUATION

To evaluate CIoTA, we used our IoT simulation testbed (Fig. 3) consisting of 48 Raspberry Pis to emulated two types of IoT devices: smart cameras and smart light. The executed attacks were either (1) code injection [21] or (2) code reuse [6]. Our experiments were carried out as follows:

1) CIoTA agents were installed on each of the Raspberry Pis, together with a symmetric key (PKI was not used in the evaluation).
2) After approximately one hour of benign operation, one of the Pis was attacked with an exploit which then executed malicious code.
3) The malicious logic behaved as a bot which attempts to connect to it's C&C once every minute.

All agents were initialized with the following parameters:

- $T$ (**Broadcast interval**): one minute
- $L$ (**Block size**): 20
- $p_c$ (**Combined model, element-wise consensus**): 75%
- $\alpha$ (**Verification distance**): 0.05
- $p_{thr}$ (**Anomaly score threshold**): 0.025

### A. Performance of the Combined Model

To evaluate the performance of a combined model, we measured its anomaly score before and after a code injection attack. The probabilities were averaged over windows of 1000 observations. Fig. 4 shows that by joining more models together, a stronger consensus is achieved much faster than using a single model. With $L = 20$, we can see that the malware's activity can be easily detected with zero false positives. Thus, the knowledge of the many is captured in the combined model, which assists each agent in distinguishing between new benign events and malicious activities. The results were similar for both exploits and use cases.

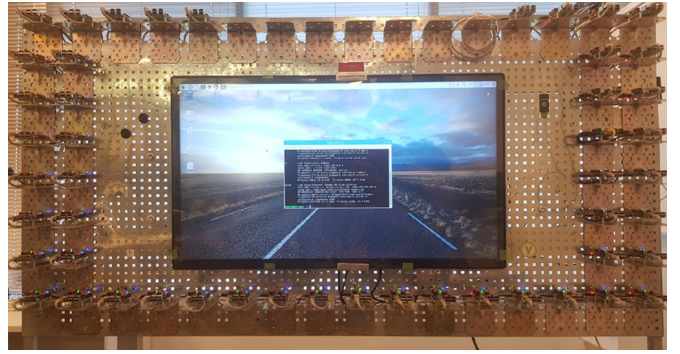

Fig. 3. IoT simulation testbed consisting of 48 Raspberry Pis

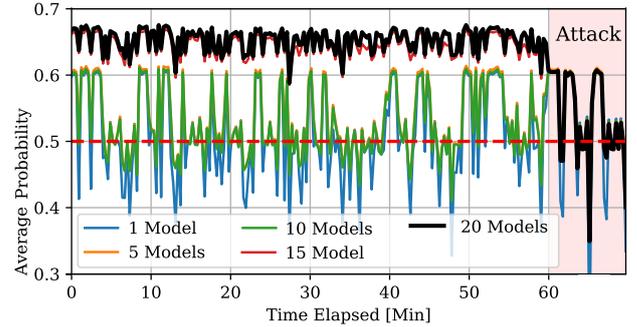

Fig. 4. The anomaly scores of the combined model $M_L$ when created from various numbers of different agent's models. The red area marks the code injection attack, followed by the malicious code's execution.

### B. Attacks Against CIoTA

To evaluate how well CIoTA performs against adversarial attacks, we examine the case where attacks were performed when CIoTA is the most vulnerable (at CIoTA's deployment, before any device has shared its chain). During this time, we attacked one or more devices with an exploit, followed by some malicious code execution. Fig. 5 plots the heat maps of the linear distance between a random agent's $M_L$ and the combined model in the *partial block* (with and without the infected model). The figure shows that if the malicious code is not similar to the original application, then other agents will reject the *partial block* due to discrepancies in the combined model (Algorithm 2).

We also investigated the case where the attacker's malicious code is similar to the application's control-flow (similar transitions in memory). As a result, all anomaly scores were borderline but did not exceed the threshold (no alerts were raised), and the modified $M_L$ was propagated to the other agents via the *partial block* of the chain. In this case, we observed that the other agent's were still capable of detecting the attack although they were using the infected model. This is because Algorithm 1 filter new behaviors which did not achieve consensus. Fig. 6 illustrates this observation in the camera use case. The figure shows that even if an attacker compromises several models, the attacker must compromise at least $p_c$ of the models in the *partial block* to evade detection.

### C. Application's Overhead

Since CIoTA is designed for resource limited devices, we measured the CPU utilization and memory consumption during runtime. We note that our code was not optimized (we used the boost and OpenSSL libraries). Therefore, our results provide an upper limit for CIoTA's overhead. The CPU utilization, memory consumption, and executable size of the agent was 6.5%, 60KB, and 260KB respectively.

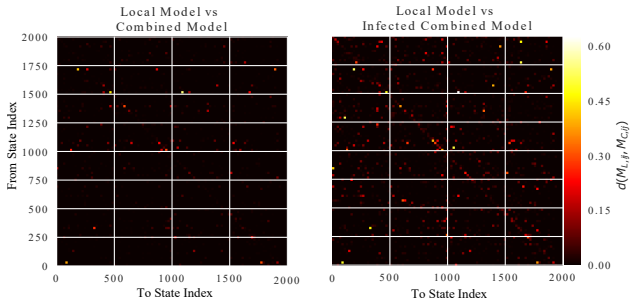

Fig. 5. Heat maps of the linear distance of a local model from a combined model. Left: clean combined model. Right: infected combined model.

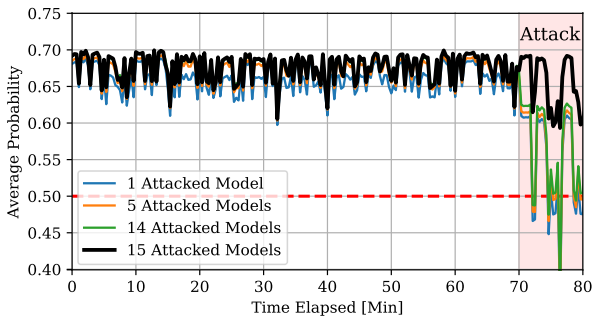

Fig. 6. The anomaly scores of the combined model $M$ using the latest block $B$, where various numbers of the models in $B$ have been infected (attacked).

## VII. CONCLUSION

In this paper, we introduced CIoTA: a blockchain-based solution for collaborative anomaly detection among a large number of IoT devices. CIoTA continuously train an anomaly detection model while remaining robust to adversarial attacks. By utilizing the wisdom of the many, CIoTA can also differentiate between rare benign events and malicious activities. A disadvantage of CIoTA is that a separate chain must be published per IoT model/firmware. Therefore CIoTA, in its present form, is most applicable to large industrial settings and smart cities. In the future, we plan to extend CIoTA to support several frameworks and improve its detection capability, e.g., by investigating API flows as opposed to the lower level control-flows.